\documentclass[12pt]{article}
\usepackage{amsmath}
\usepackage{graphicx, color}
\usepackage[sort&compress,square,numbers]{natbib}
\usepackage{url} % not crucial - just used below for the URL 

\newcommand{\blind}{0}

\begin{document}

\def\spacingset#1{\renewcommand{\baselinestretch}%
{#1}\small\normalsize} \spacingset{1}

%%%%%%%%%%%%%%%%%%%%%%%%%%%%%%%%%%%%%%%%%%%%%%%%%%%%%%%%%%%%%%%%%%%%%%%%%%%%%%

\if0\blind
{
  \title{\bf Design Principles for Data Analysis}
  \author{Lucy D'Agostino McGowan\\
    Department of Mathematics and Statistics, Wake Forest University\\
    and \\
    Roger D. Peng \\
   Department of Biostatistics, Johns Hopkins Bloomberg School of Public Health \\ 
    and \\ 
    Stephanie C. Hicks \\
   Department of Biostatistics, Johns Hopkins Bloomberg School of Public Health}
  \maketitle
} \fi

\if1\blind
{
  \bigskip
  \bigskip
  \bigskip
  \begin{center}
    {\LARGE\bf Design Principles for Data Analysis}
\end{center}
  \medskip
} \fi

\bigskip
\begin{abstract} % 200 or fewer words
The data science revolution has led to an increased interest in the practice of data analysis. While much has been written about statistical thinking, a complementary form of thinking that appears in the practice of data analysis is design thinking -- the problem-solving process to understand the people for whom a product is being designed. For a given problem, there can be significant or subtle differences in how a data analyst (or \textit{producer} of a data analysis) constructs, creates, or designs a data analysis, including differences in the choice of methods, tooling, and workflow. These choices can affect the data analysis products themselves and the experience of the \textit{consumer} of the data analysis. Therefore, the role of a producer can be thought of as designing the data analysis with a set of design principles. Here, we introduce \textit{design principles for data analysis} and describe how they can be mapped to data analyses in a quantitative, objective and informative manner. We also provide empirical evidence of variation of principles within and between both producers and consumers of data analyses. Our work leads to two insights: it suggests a formal mechanism to describe data analyses based on the design principles for data analysis, and it provides a framework to teach students how to build data analyses using formal design principles.
\end{abstract}

\noindent%
\vfill

\newpage
\spacingset{1.5} % DON'T change the spacing!
\section{Introduction}
\label{sec:intro}

The data revolution has led to an increased interest in the practice of \textit{data analysis} \cite{tuke:1962, tukeywilk1996, box1976, wild1994, chatfield1995, wildpfannkuch1999}. In the practice of data analysis, one often uses \textbf{statistical thinking} \cite{wildpfannkuch1999}, namely the vague but intuitive process of aiming to accurately describe or understand uncertainties in a complex world using foundations from mathematics, statistics, computer science, psychology, and other fields of study \cite{Snee1990statthinking,  Chance2002statthinking, Poldrack2021book}. This often manifests where, for a given question or decision that needs to be made, a data analyst (or \textit{producer} of a data analysis) makes analytic choices, such as which methods, algorithms, computational tools, languages or workflows to use in a data analysis \cite{grol:wick:2014, Donoho2017} that most accurately capture or describe a complex world. For example, a data analysis can consist of simply calculating the sample mean for a given set of observed data. Alternatively, the producer may choose to calculate a sample median if they suspect there are outliers in the observed data. A data analysis can also be more complicated consisting of, for example, importing, cleaning, transforming and modeling data with a goal to build a machine learning algorithm to decide which product a company should sell. 

In contrast to the goal of describing a complex world accurately through statistical thinking, alternative, but complementary, forms of thinking also appear in the practice of data analysis, including \textbf{design thinking} \cite{Cross2011design, Parker2017, Woods2019designthinking, nolis2020build}. This iterative, solutions-based, problem-solving process aims to understand and deeply empathize with the people for whom a product is being designed for \cite{Cross2011design}. A common feature of design thinking is to employ divergent thinking, or the process of identifying and exploring many solutions (possible or impossible) \cite{Cross2021divergent}. This is in contrast to convergent thinking, often used in statistical thinking, where the choices, which can be influenced by factors outside the control of the producer such as time or budget constraints \cite{peng2018-resources} or the availability of appropriate data, are narrowed down to a final solution that is most accurate or correct for given a problem. In the practice of data analysis, one way divergent thinking often manifests is, for a given question or decision that needs to be made, a producer can explore the space of (i) how information from the data is extracted, summarized, and presented \cite{cook2007, Parker2017}, (ii) the degree to which evidence in the data is reported or is convincing or the degree to which alternative methods or approaches are considered \cite{wildpfannkuch1999, Breiman2001}, and (iii) how reproducible the data analysis is \cite{Knuth1984}. Ultimately, these design choices for a given data analysis impact the final product that is produced \cite{nolis2020build}. For example, a producer of a data analysis can prioritize exhaustively checking a set of assumptions of a specific method instead of making a more modest effort. While this prioritization often leads to a longer data analysis, it can also lead to different results (or a different interpretation of results) if the assumptions of the method are found to not be supported by the data. Previous empirical studies have found that even when using the same data to investigate the same question, there can be significant variation in how producers build data analyses, which has been shown to influence the results of the analysis \cite{Silberzahn2018}. 

These prioritized factors or characteristics can not only induce variation in the data analyses themselves, but also can affect a \textit{consumer} (or stakeholder \citep{nolis2020build}) of the data analysis. Using the same example as above, when a producer prioritizes exhaustively checking a set of assumptions of a specific method, the experience of a consumer of the data analysis (who was expecting an exhaustive analysis) might also be changed from being less confident to more confident as the degree of exhaustively checking the assumptions increases. Alternatively, if a producer makes a design choice to summarize the results from a data analysis with only tables, then a consumer (who was expecting summaries with plots) might not understand the results without data visualizations and therefore be skeptical of any results. 

We refer to these prioritized factors or characteristics that are relevant to the data analysis, as a whole or individual components, as \textit{design principles} for data analysis. Broadly, when building a data analysis, the role of a producer can be thought of as designing the data analysis with a set of data analytic principles to serve a larger purpose, such as to be able to extract meaningful information, answer an original question, support decision-making, or address the needs or expectations of data analysis consumers. Similar to principles of art or music \cite{Lambert2014-music}, the design principles for data analysis are not meant to be used to evaluate the quality of a data analysis, but rather they are meant to be objective characteristics about the data analysis that can be used to induce or describe variation between data analyses. 

Our primary focus in this manuscript is to (i) introduce the data analytic design principles (Section~\ref{sec-principles}), (ii) describe an example in the classroom for how the design principles can be mapped to data analyses in a quantitative, objective and informative manner and demonstrate empirical evidence of variation in principles within and between both producers and consumers of data analyses (Section~\ref{sec-results}). In the Discussion (Section~\ref{sec-discussion}), we discuss how these data analytic design principles can be implemented in practice: how the design principles might be used to evaluate the quality of a data analyses, or how the design principles can be used in the classroom by practitioner-instructors \cite{KrossGuo2019} to build data analyses. 

\section{Design principles for data analysis}
\label{sec-principles}

The design principles for data analysis are prioritized qualities or characteristics that are relevant to the analysis and can be objectively observed or measured. Driven by statistical thinking and design thinking, a data analyst can use these principles to guide the choice of which data analytic \textit{elements} to use, such as code, code comments, data visualization, non-data visualization, narrative text, summary statistics, tables, and statistical models or computational algorithms~\cite{Breiman2001}, to build a data analysis. A data analysis can be \textit{scored} based on how well it adheres to each of these principles. The scoring is not meant to convey a value judgment with respect to the overall quality of the data analysis. Rather, the requirement is that multiple people viewing an analysis could reasonably agree on the fact that an analysis gives high or low score to certain principles. Value judgments may be overlaid on to an analysis by the consumer based on how different principles are scored, but we do not consider such judgments universal characteristics. Next, we describe six principles that we hypothesize are informative for characterizing variation between data analyses.

\vspace{0.5em}\noindent\textbf{Data Matching}. Data analyses with high \textit{data matching} have data readily measured or available to the producer that directly match the data needed to investigate a question (Figure~\ref{fig-data-matching}). In contrast, a question may concern quantities that cannot be directly measured or are not available to the producer. In this case, data matched to the question may be surrogates for covariates that measure the underlying data phenomena. While we consider the main question and the data to be contextual inputs to the data analysis, we consider this a design principle of data analysis because the producer selects methods, tooling, or workflows that are used to investigate the question, which depend on how well the data are matched. If the data are poorly matched, the producer will not only need to investigate the main question with one set of methods, but also will need to use additional methods that describe how well the surrogate data are related to the underlying data phenomena. 

It is important to note that questions can be more or less specific, which will impose strong or weak constraints on the range of data matching to the question. Highly specific questions tend to induce strong constraints to investigate with which what methods, tooling, or workflows are used. Less specific questions emit a large range of potential data to investigate the question. Data that can be readily measured or are available to the producer to directly address a specific question results in high data matching, but depending on the problem specificity, can result in a narrow or broad set of data to consider. 

\vspace{0.5em}\noindent\textbf{Exhaustive}. An analysis is \textit{exhaustive} if specific questions are addressed using multiple, complementary methods, tooling or workflows (Figure~\ref{fig-exhaustive}). For example, using a $2\times 2$ table, a scatter plot, and a correlation coefficient are three different tools that could be employed to investigate whether two predictors are correlated. Analyses that are exhaustive use multiple, complementary tools or methods to address the same question, knowing that each given tool reveals some aspects of the data but obscures other aspects. As a result, the combination of tools and methods used may provide a more complete picture of the evidence in the data than any single tool would.

\vspace{0.5em}\noindent\textbf{Skeptical}. An analysis is \textit{skeptical} if multiple, related questions are considered using the same data (Figure~\ref{fig-skeptical}). Analyses, to varying extents, consider alternative explanations of observed phenomena and evaluate the consistency of the data with these alternative explanations. Analyses that do not consider alternate explanations have no skepticism. For example, to examine the relationship between a predictor X and an outcome Y, an analysis may choose to show results from different models containing different sets of predictors that might potentially confound that relationship. Each of these different models represents a different, but related, question about the X-Y relationship. A separate question that arises is whether the configuration of alternative explanations are relevant to the problem at hand. However, often that question can only be resolved using contextual information that is outside the data. 

The need for more or less skepticism in a data analysis is typically governed by outside circumstances and the context in which the analysis sits. Analyses that may have large impacts or result in significant monetary costs will typically be subject to detailed scrutiny. In July 2000, the Health Effects Institute (HEI) published a reanalysis of the Harvard Six Cities Study, a seminal air pollution study that showed significant associations between air pollution and mortality. Due to the potential regulatory impact of the study, HEI commissioned an independent set of investigators to reproduce the findings and conduct a series of sensitivity analyses~\citep{krew:burn:gold:hoov:2000}. The result was a nearly 300 page volume where the data and findings were subject to intense skepticism and every alternative hypothesis was examined. 

There are other instances when skepticism in the form of alternate explanations is not warranted in the analysis. For example, with an explicitly planned and rigorously-conducted clinical trial, the reported analysis will typically reflect only what was pre-specified in the trial protocol. Other analyses may be presented in a paper but they will be explicitly labeled as secondary. For example, in a large clinical trial studying the effect of a pest management intervention on asthma outcomes~\cite{matsui2017effect}, the reported analysis is ultimately a simple comparison of asthma symptoms in two groups. Some other secondary analyses are presented, but they do not directly address the primary question. Such an analysis is acceptable here due to the strict pre-specification of the analysis and due to the standards and practices that the community has developed regarding the reporting of clinical trials.

\vspace{0.5em}\noindent\textbf{Second-Order}. An analysis is \textit{second-order} if it includes methods, tooling or workflows that do not directly address the primary question, but give important context or supporting information to the analysis (Figure~\ref{fig-second-order}). Any given analysis will contain, for example, data visualizations that directly contribute to the results or conclusions, as well as some other data visualizations that provide background or context or are needed for other reasons (Figure~\ref{fig-data-matching}). Second-order analyses contain more of these background/contextual elements in the analysis, for better or for worse. For example, in presenting an analysis of data collected from a new type of machine, one may include details of who manufactured the machine, why it was built, or how it operates. Often, in studies where data are collected in the field, such as in people's homes, field workers can relay important details about the circumstances under which the data were collected. In both examples, these details may be of interest and provide useful background, but they may not directly influence the analysis itself. Rather, they may play a role in helping a consumer interpret the results and evaluate the strength of the evidence.

\vspace{0.5em}\noindent\textbf{Clarity}. Analyses with \textit{clarity} summarize or visualize data in a way that is influential in explaining how the underlying data phenomena or data-generation process connects to any key output, results, or conclusions (Figure~\ref{fig-transparent}). Clarity could be demonstrated by simply including one data visualization, or it could consist of multiple data visualizations. While the totality of an analysis may be complex and involve a long sequence of steps, analyses with clarity summarize or visualize key pieces of evidence in the data that explain the most ``variation'' or are most influential to understanding the key results or conclusion. One aspect of exhibiting clarity is showing the approximate mechanism by which the data inform the results or conclusion.

\vspace{0.5em}\noindent\textbf{Reproducible}. An analysis is \textit{reproducible} if someone who is not the original producer can take the published code and data and compute the same results as the original producer (Figure~\ref{fig-reproducible}). Critical to reproducibility is the availability of a stable form for both the dataset and the analytic code. For example, an analysis might consist of interactively calculating a sample mean for a given set of observations in the console of a programming language. In this case, as there is no stable code because the analysis was performed interactively, it is not possible for another person to reproduce the analysis. These types of analyses that are not deliberately recorded happen frequently and do not necessarily imply a negative quality about the analysis. Rather such analyses are simply not reproducible. In contrast, analyses that integrate literate programming \cite{Knuth1984} in an analytic compendium are more reproducible \cite{Vassilev2016-literate}. Another consideration is that it may not be possible for businesses, such as those in the finance industry, to make available entire analytic compendia for proprietary or financial reasons. In contrast, analytic compendia that are integrated as part of the analytic product or analytic presentation are by definition more reproducible. Finally, much has been written about reproducibility and its inherent importance in science, so we do not repeat that here~\cite{peng:2011}. We simply add that reproducibility (or lack thereof) is usually easily verified and is not dependent on the characteristics of the consumer of the analysis. Reproducibility also speaks to the coherence of the workflow in the analysis in that the workflow should show how the data are transformed to eventually become results.

\section{Results}
\label{sec-results}

In this section, we describe two case studies of how these principles can be applied in the classroom (one at Wake Forest University and one at Johns Hopkins University) and mapped to data analyses in a quantitative, objective and informative manner. We give empirical evidence for variation in principles within and between both producers and consumers of data analyses. 

\subsection{Analysis}

Our approach to data analysis was exploratory with the goal of summarizing the data. We examine the between person and within person variation across principles. We first examine how this may vary for producers of data analysis via the Wake Forest data. Then we examine variation in responses for consumers of data analysis via the Johns Hopkins data.

\subsubsection{Wake Forest University Data}

Participants consist of 54 students enrolled in a Statistical Models course at Wake Forest University. This course is intended for students who have had at least one university-level statistics course. The study was approved by the Wake Forest University Institutional Review Board (IRB00023932). 

Participants were taught the 6 design principles of data analysis. Throughout the course, they were given 8 data analysis assignments. On each of the 8 assignments, they were asked to rank the analysis they completed from 1 to 10 across each of the principles, with one indicating that the analysis did not adhere to the principle, and 10 indicating that it did. We also collected data on the participants' current major.

\begin{figure}[tbh]
    \centering
    \includegraphics[width=6in]{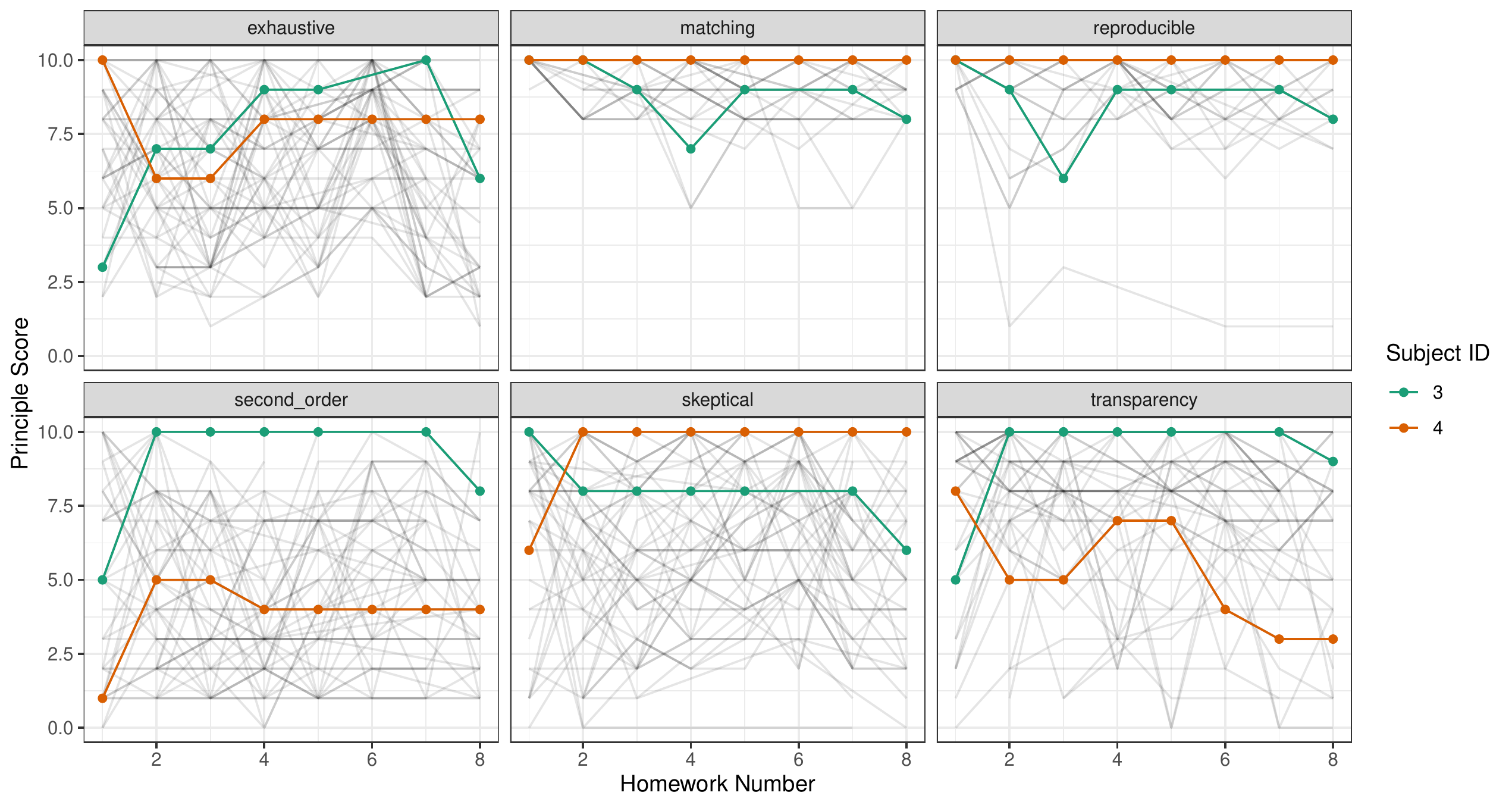}
    \caption{Between and within person variation of principles across assignments.}
    \label{fig:withinperson}
\end{figure}

\subsubsection{Variation in principles among producers of data analyses}
\label{sec-variation-1}

\begin{figure}
    \centering
    \includegraphics[width=5in]{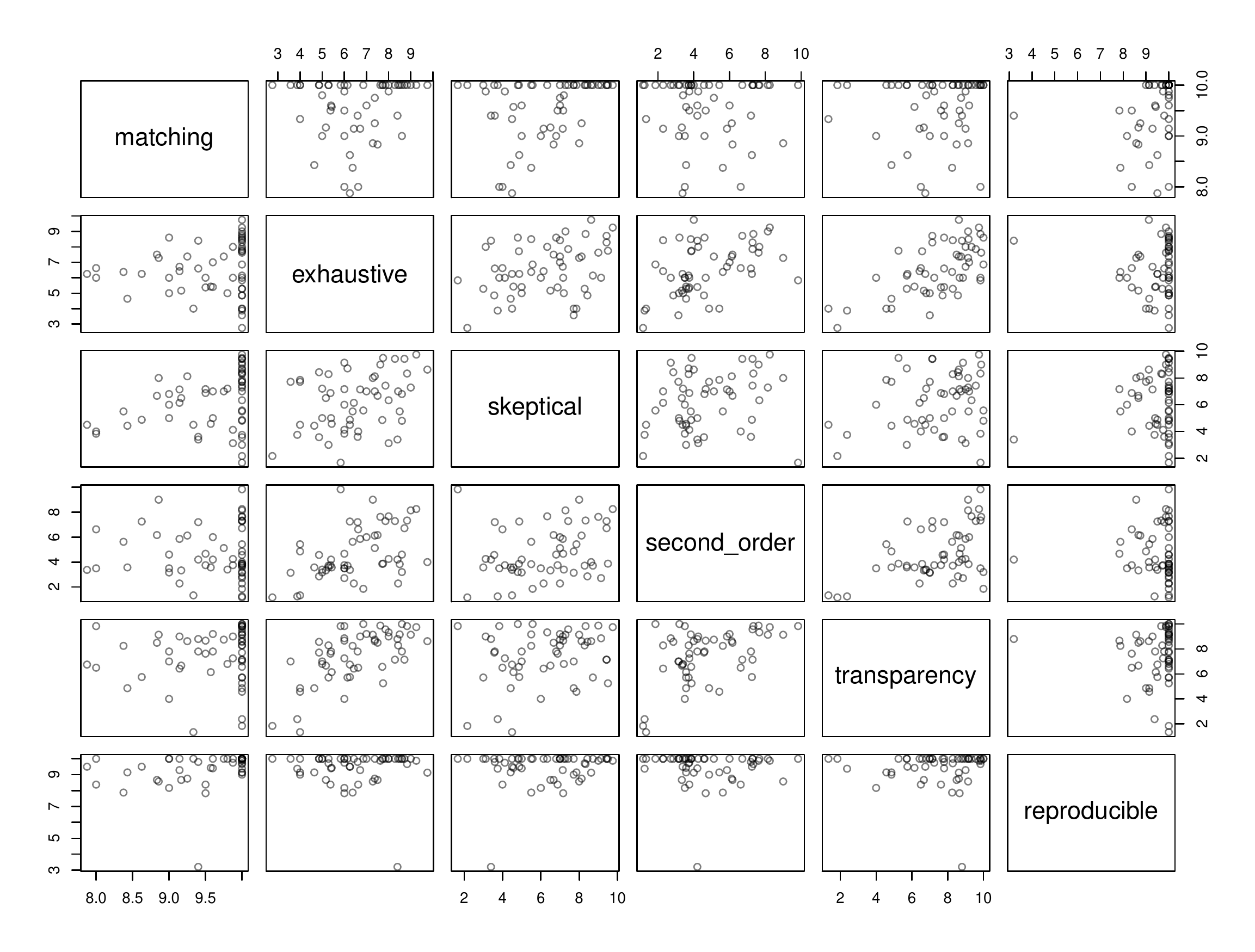}
    \caption{Pairwise correlation between principles.}
    \label{fig:correlation}
\end{figure}

\begin{figure}
    \centering
    \includegraphics[width=5in]{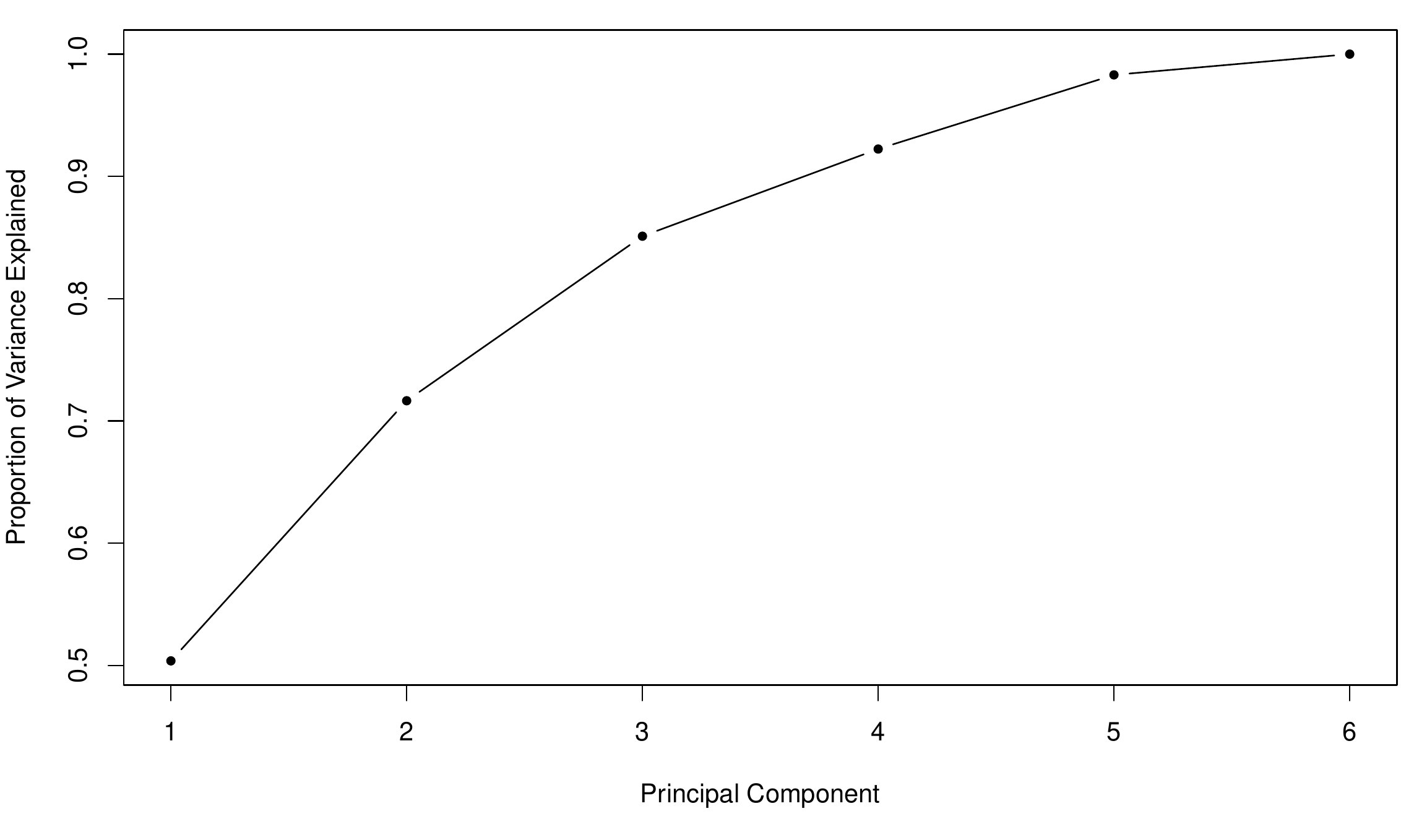}
    \caption{Cumulative proportion of variance explained by principal components.}
    \label{fig:cumulativevar}
\end{figure}

We consider the participants from Wake Forest University as \textit{producers}. Among these data analysis producers, we see variation both between and within persons (Figures \ref{fig:withinperson}, \ref{fig:correlation}). Figure \ref{fig:withinperson} shows each individual's score of  the six principles across the 8 analyses. Two ``profiles'' are selected for demonstration purposes to illustrate both the variability in scores within a given individual as well as the variability between individuals. Figure \ref{fig:correlation} shows the pairwise correlation between principles; while some principle scores are more highly correlated than others, there appears to consistently be variability, indicating that these principle scores measure different underlying characteristics of a data analysis. This can be better visualized by Figure \ref{fig:cumulativevar}. The cumulative proportion of variance explained by principle components illustrates that not all principles are loaded in a single component, again suggesting that there is additional information added across the six principles.

\begin{figure}
    \centering
    \includegraphics[width=5.5in]{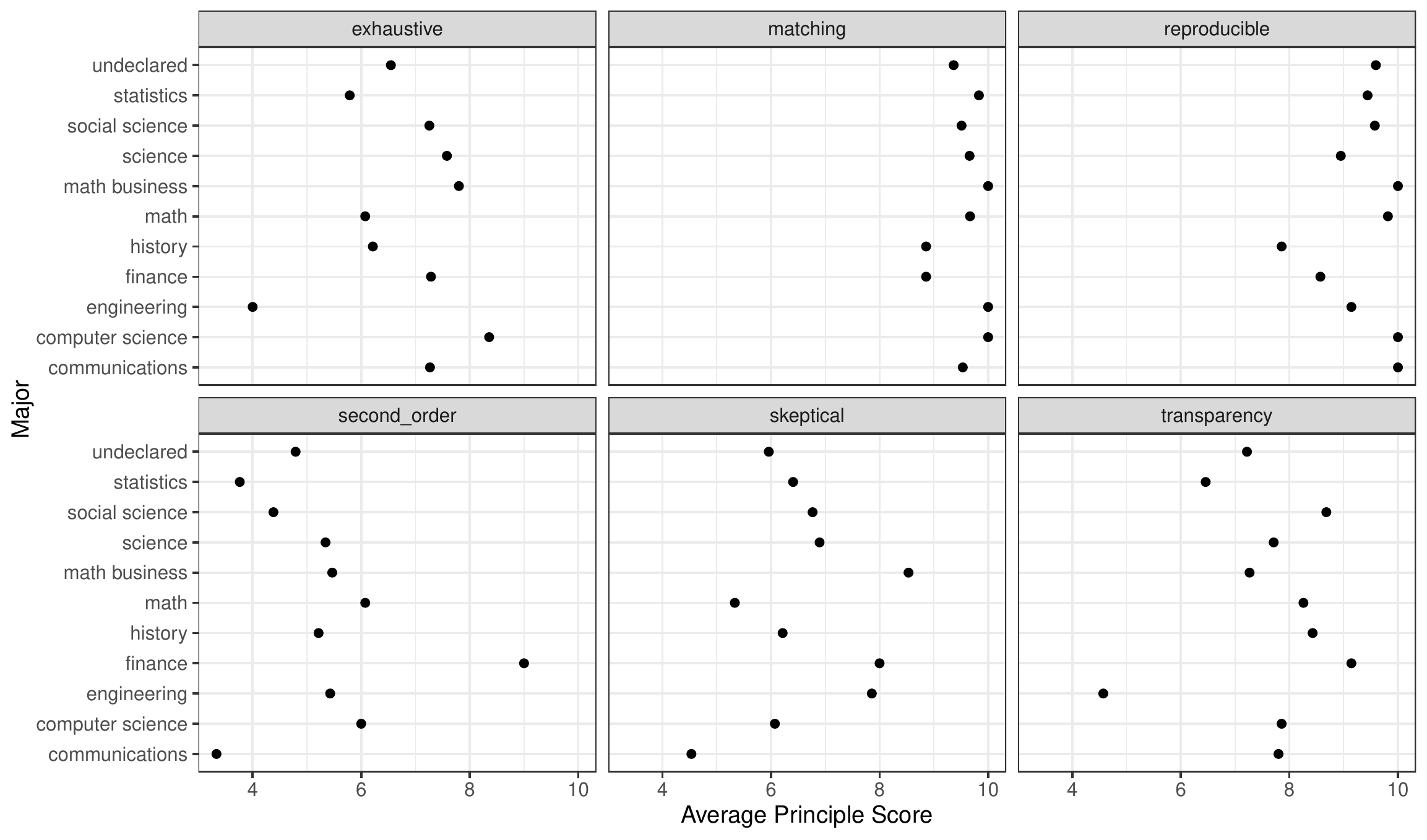}
    \caption{Average principle scores by declared major.}
    \label{fig:majors}
\end{figure}

In Figure \ref{fig:withinperson}, we observed substantial between person variability in some principles. It is possible that this is partially explained by a producer's baseline characteristics.  As an illustrative example, we examine the average principle score by the producers' declared major (Figure \ref{fig:majors}). Some principles exhibit more variability than others, suggesting that major may be a driving factor. For example, finance majors report higher scores for the \textit{second order} principle and engineering majors report lower scores for \textit{exhaustive} and \textit{clarity} principles.

\subsubsection{Johns Hopkins University Data}

%JHU description
Data from Johns Hopkins University were collected from 15 students enrolled in a course offered in the Department at Johns Hopkins titled Advanced Data Science. For a homework assignment, students were asked to evaluate a data analysis completed by two separate authors using a score of 1 to 10 for each of the principles described in Section~\ref{sec-principles}. The data analyses consisted of analyses of natural disasters in the United States and their economic impact. Each analysis was done by a different person, but the datasets and question addressed were the same. The students were given the output of the data analysis, but were not given the data and were not asked to analyze the data themselves.

The students in the course were also asked to take two different perspectives when evaluating the principles. The first was the perspective of an EPA policy official in need of evidence for the impact of natural disasters. The second perspective was that of a homebuyer interested in purchasing a home in an area susceptible to natural disasters. 

The dataset consists of two sets of scores from each student, one set from the EPA perspective and one set from the homebuyer perspective. This study was approved by the Institutional Review Board of the Johns Hopkins Bloomberg School of Public Health (IRB 00012419).

\subsubsection{Variation in principles among consumers of data analyses}
\label{sec-variation-2}

\begin{figure}[ht!]
  \centering\includegraphics[width=.75\textwidth]{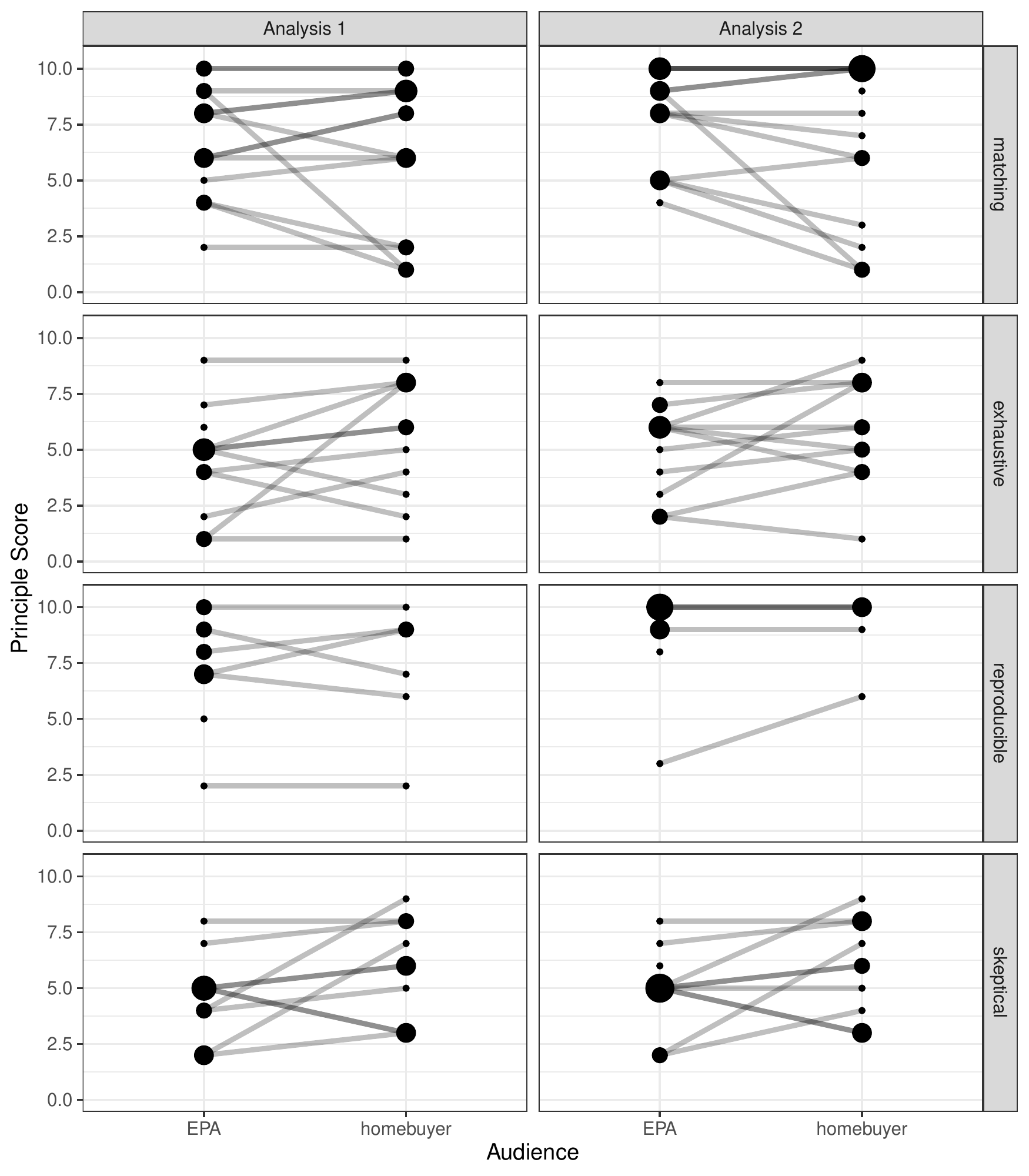}
  \caption{Variation in principle scores by consumer audience type. The size of the circles is proportional to the number of individuals assigning that score value.}
  \label{fig:jhudata}
\end{figure}

We consider the participants from Johns Hopkins University as \textit{consumers}. Among data analysis consumers, we see variation between audiences across two analyses (Figure \ref{fig:jhudata}). Students were asked to assess two analyses, each from the perspectives of two consumers: the Environmental Protection Agency (EPA) and a potential home-buyer who is interested in purchasing a house along the coast in the state of Florida. While the direction of the difference in scores differs between individuals (i.e. some have higher scores for particular principles when viewing the analysis through the EPA's lens compared to a home-buyer's lens and vice versa), consumers did differ in their scores, as evidenced by the non-zero slope of most lines in Figure \ref{fig:jhudata}. For example, the \textit{matching} scores for home-buyer were generally lower than for EPA, with an average of 6.6 for home-buyer versus 7.3 for EPA. Scores for \textit{skeptical} were higher for home-buyer relative to EPA, with an average of 5.8 for home-buyer and 4.8 for EPA.  

Some principles had more variability than others, for example \textit{matching}, \textit{exhaustive}, and \textit{skeptical} seem to have more variability compared to \textit{reproducible}. Additionally, there are noticeable differences between analyses. For example across the \textit{reproducible} principle, Analysis 2 has higher scores overall compared to Analysis 1, with an average score of 8.9 for Analysis 2 and an average score of 7.4 for Analysis 1.

\label{sec-practice}

\section{Discussion}
\label{sec-discussion}

% In the Discussion (Section~\ref{sec-discussion}), we discuss how these data analytic design principles can be implemented in practice: how the design principles might be used to evaluate the quality of a data analyses, or how the design principles can be used in the classroom by practitioner-instructors \cite{KrossGuo2019} to build data analyses. 

In this paper we have introduced a framework of design principles that can be used for constructing data analyses and demonstrated the implementation of this framework in two classroom settings aimed at teaching data analysis. In our analysis of the classroom data, we found that the principles defined here appear to consistently measure underlying quantities that are reasonably uncorrelated with each other. Furthermore, it appeared that the design principles could be used to examine differences between groups, including different majors~(Figure~\ref{fig:majors}) or consumer audiences~(Figure~\ref{fig:jhudata}).

One significant consequence of using design thinking concepts in data analysis is that it allows for the explicit separation of producers and consumers of a data analysis. Traditional descriptions of statistical thinking generally conceive of a single analyst building data analyses and obtaining feedback on their approach from the data. While the notion of a consumer for that analysis may be embedded in the idea of statistical thinking, it is often not well-specified. The benefit of conceptually separating producers from consumers is that such a separation serves to demonstrate potential differences in priorities between the two groups. 

\begin{figure}
    \centering
    \includegraphics[width=.75\textwidth]{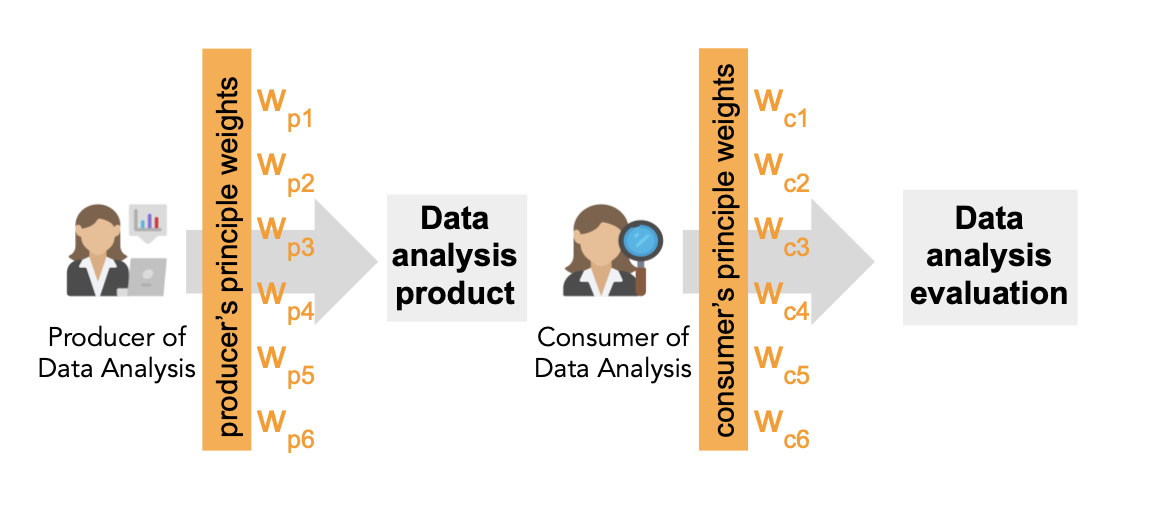}
    \caption{Conceptual framework to describe how producers and consumers have principle weights that inform how they produce or characterize a data analysis. The weights for the producer are represented as $w_{pi}$ for each of the six principles, and the weights for the consumers are represented as $w_{ci}$ for each of the principles.}
    \label{fig:product_evaluation}
\end{figure}

In general, consumers of data analyses will have certain expectations for what they see and data analysts (producers) can construct an analysis that either meets those expectations or not. One possible way to quantify a consumer's expectations for a data analysis is to assign \textit{a priori} weights to each of the design principles described here. The distance between a consumer's weights on these principles and the scores assigned to the realized data analysis could indicate the extent to which the analysis meets the consumer's expectations. Producers may also assign \textit{a priori} weights to the different principles that can guide the construction of an analysis. If a producer's and consumer's weights are known to be substantially different from each other, then this would be an \textit{a priori} indication that the analysis may not meet the consumer's expectations~(Figure~\ref{fig:product_evaluation}). In such a situation, it may be valuable for the producer and the consumer to come to an agreement over the weighting of the principles before time is spent doing the analysis. Here, the design principles provide a formal articulation of how producer and consumer can agree (or disagree) on the ultimate outcome.

The design thinking perspective on data analysis also has useful consequences for teaching data analysis in the classroom, where it is valuable have a formal way to describe what makes data analyses differ from each other and why one type of analysis might be preferable in some circumstances to another type of analysis. In particular, in teaching about the divergent thinking phase of data analysis, it is common to encourage students to take different approaches to addressing a data analytic question. However, we often lack a formal basis for characterizing these different approaches for students. The data analytic design principles provide one way to separate different approaches and to guide students to explore various approaches to problem solving.

Concepts from design thinking can serve as important complements to the traditional notion of statistical thinking. Together, these two forms of thinking provide a more complete road map for developing useful data analyses and present new ways to teach data analyses to novices. The formal specification of design principles for data analysis and how they may guide data analysis construction offers a rationale for negotiating qualities of a data analysis between producer and consumer before embarking on substantial data analytic work. An area for possible future work includes measuring to what extent manipulating the weighting of these principles can improve the quality of data analysis.

%We have demonstrated the Data analysis consumers will have certain preferences or priorities about which principles should be prioritized in a data analysis. One way to characterize the success of an analysis is to look at the distance between the objective score of the analysis and what individual weights or preferences are for these principles. Similarly, people who produce analyses weight these principles in a certain way and that will guide their production of an analysis. If consumers and producers are to agree as to whether an analysis achieves its goals, they will likely need to have similar weights on these principles a priori.

%The scoring of an analysis as we have demonstrated here could be one way to measure how far an analysis is from achieving its goals by comparing the scores to the weights assigned to each principle by a consumer. Different consumers with different weights may have different opinions about whether a given analysis has achieved its goals. Figure \ref{fig:product_evaluation} describes the conceptual framework under which producers and consumers have principle weights that inform how they produce or characterize a given analysis. The producer's weights will likely influence the scores assigned to the analysis product, and the consumer's weights will determine how the product is viewed.

% References
\bibliographystyle{unsrtnat}
\bibliography{project-principles}

\clearpage 
% Supplement begins here
% Counters for Sections, references, Figs & Tables reset
% Cite a reference or main text fig/table in the usual way 
\setcounter{page}{1}

\setcounter{section}{0}
\makeatletter
\renewcommand{\thefigure}{S\@arabic\c@figure}
\makeatother

\setcounter{figure}{0}
\makeatletter
\renewcommand{\thefigure}{S\@arabic\c@figure}
\makeatother

\setcounter{table}{0}
\makeatletter
\renewcommand{\thetable}{S\@arabic\c@table}
\makeatother

\onecolumn

\bigskip
\begin{center}
{\large\bf SUPPLEMENTARY MATERIAL}
\end{center}

%\begin{description}
% \item[Title:] Supplemental figures S1-S6 (file format: PDF)
%\end{description}

\begin{figure}[ht!]
  \centering\includegraphics[width=.75\textwidth]{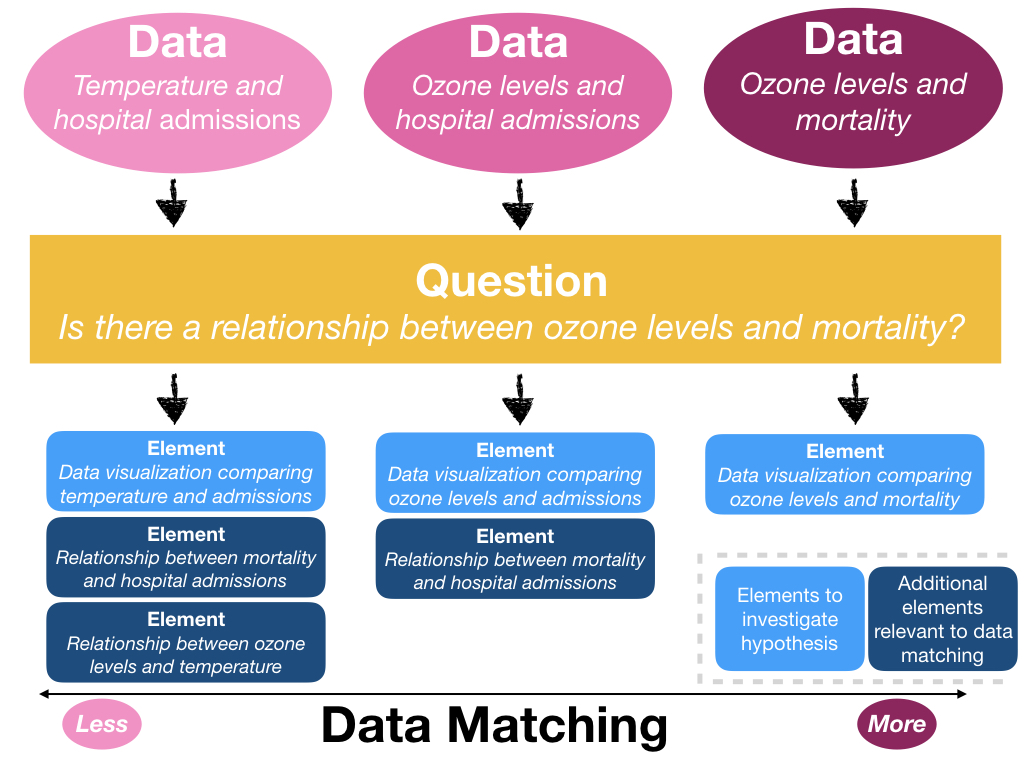}
  \caption{\textbf{The data matching principle of data analysis.} Data analyses with high data matching have data readily measured or available to the producer that directly matches the data needed to investigate a question or problem with data analytic elements. In contrast, a question may concern quantities that cannot be directly measured or are not available to the producer. In this case, data matched to the question may be surrogates or covariates to the underlying data phenomena that may need additional elements to describe how well the surrogate data is related to the underlying data phenomena to investigate the main question. }
  \label{fig-data-matching}
\end{figure}

\begin{figure}[t!]
  \centering\includegraphics[width=.75\textwidth]{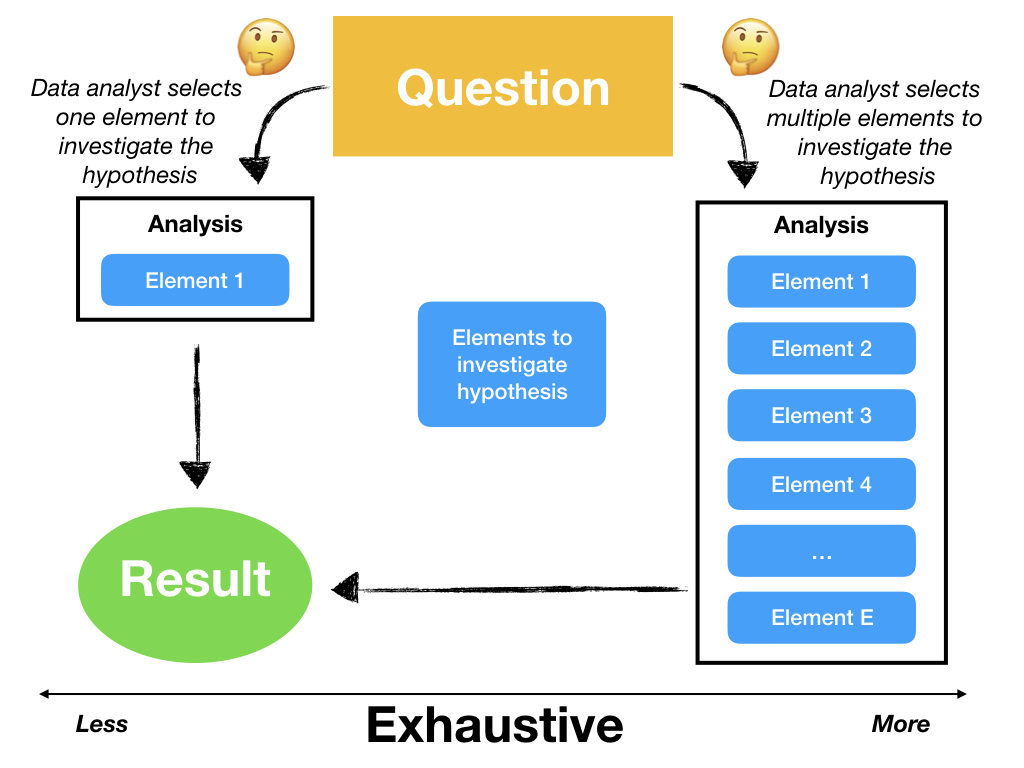}
  \caption{\textbf{The exhaustive principle of data analysis.} An analysis is exhaustive if specific questions are addressed using multiple, complementary elements. For a given question, the producer can select an element or set of complementary elements to investigate the question. The more complementary elements that are used, the more exhaustive the analysis is, which provides a more complete picture of the evidence in the data than any single element. }
  \label{fig-exhaustive}
\end{figure}

\begin{figure}[ht!]
  \centering\includegraphics[width=.75\textwidth]{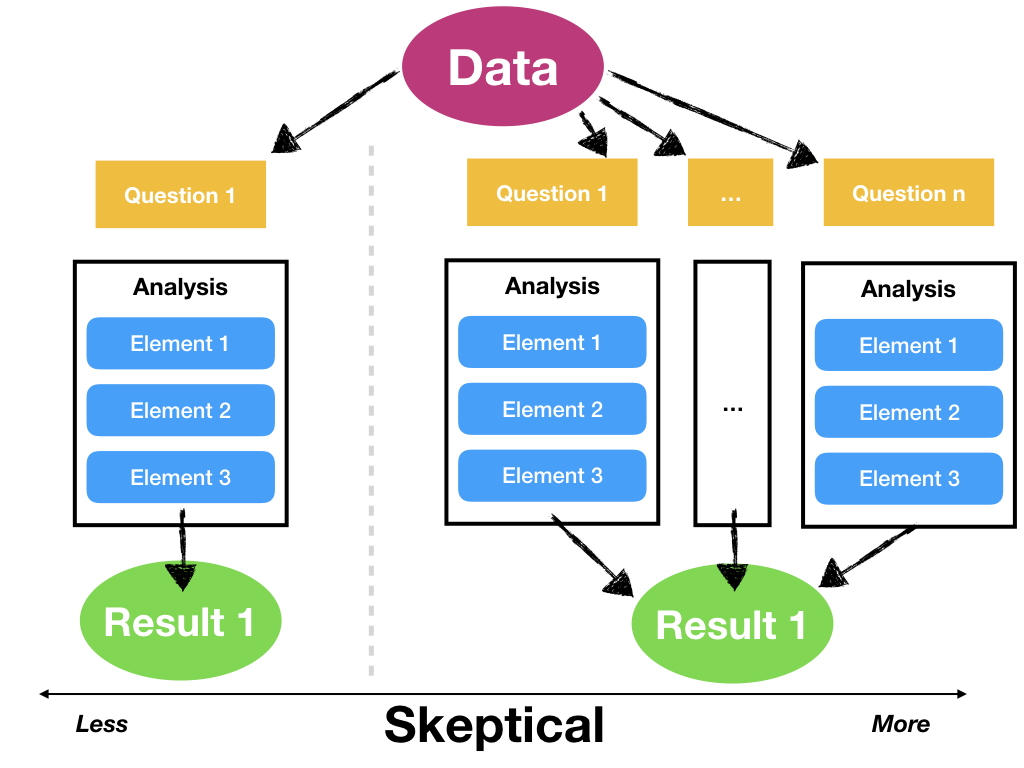}
  \caption{\textbf{The skeptical principle of data analysis.} An analysis is skeptical if multiple, related questions or alternative explanations of observed phenomena are considered using the same data and offer consistency of the data with these alternative explanations. In contrast, analyses that do not consider alternate explanations have no skepticism.}
  \label{fig-skeptical}
\end{figure}

\begin{figure}[t!]
  \centering\includegraphics[width=.75\textwidth]{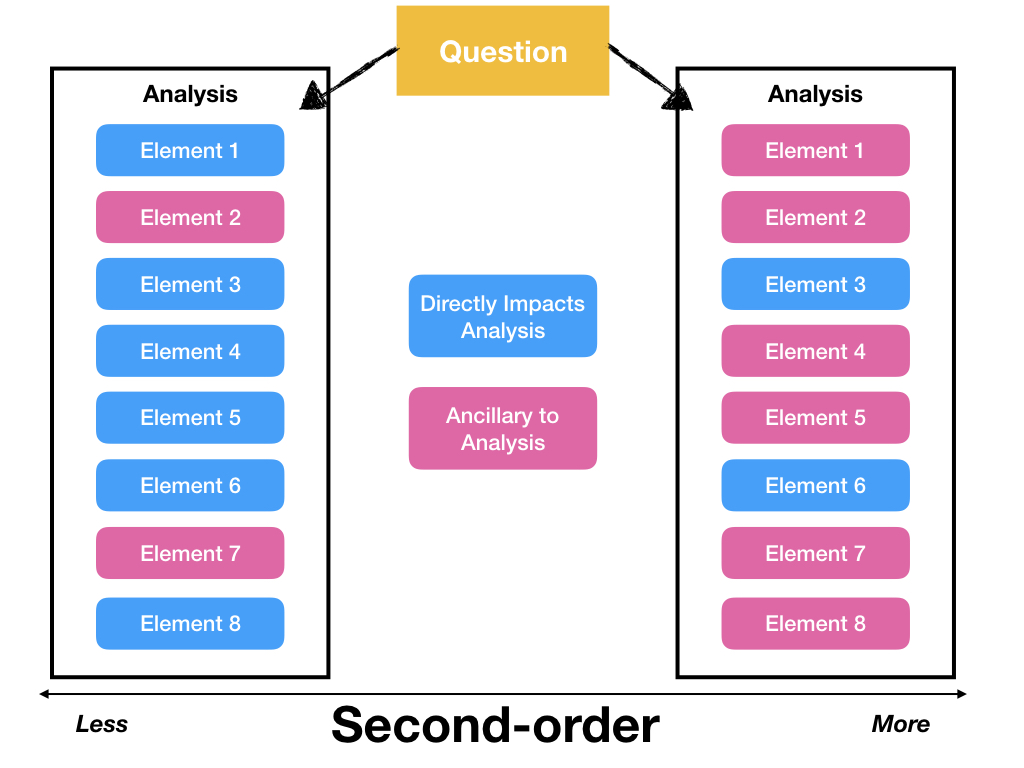}
  \caption{\textbf{The second-order principle of data analysis.} An analysis is second-order if it includes ancillary elements that do not directly address the primary question but give important context to the analysis. Examples of ancillary elements could be background information of how the data were collected, and expository explanations or analyses comparing different statistical methods or software packages. While these details may be of interest and provide useful background, they likely do not directly influence the analysis itself.
}
  \label{fig-second-order}
\end{figure}

\begin{figure}[ht!]
  \centering\includegraphics[width=.75\textwidth]{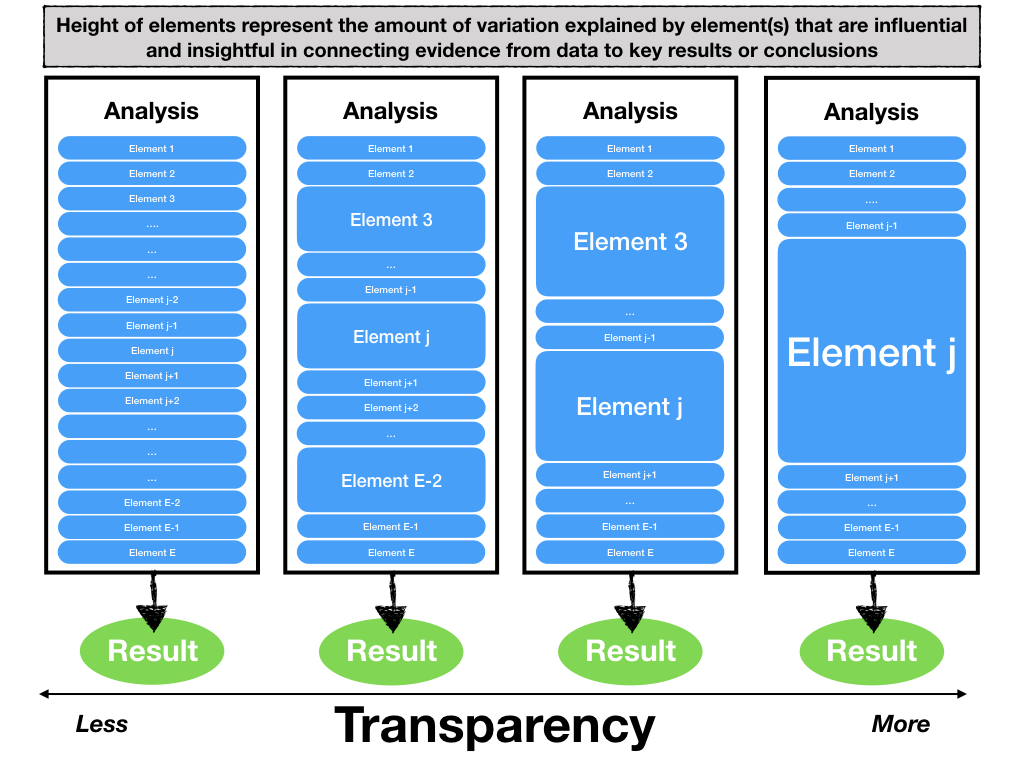}
  \caption{\textbf{The clarity principle of data analysis.} Transparent analyses present an element or set of elements summarizing or visualizing data that are influential in explaining how the underlying data phenomena or data-generation process connects to any key output, results, or conclusions. While the totality of an analysis may be complex and involve a long sequence of steps, transparent analyses extract one or a few elements from the analysis that summarize or visualize key pieces of evidence in the data that explain the most “variation” or are most influential to understanding the key results or conclusion. }
  \label{fig-transparent}
\end{figure}

\begin{figure}[ht!]
  \centering\includegraphics[width=.75\textwidth]{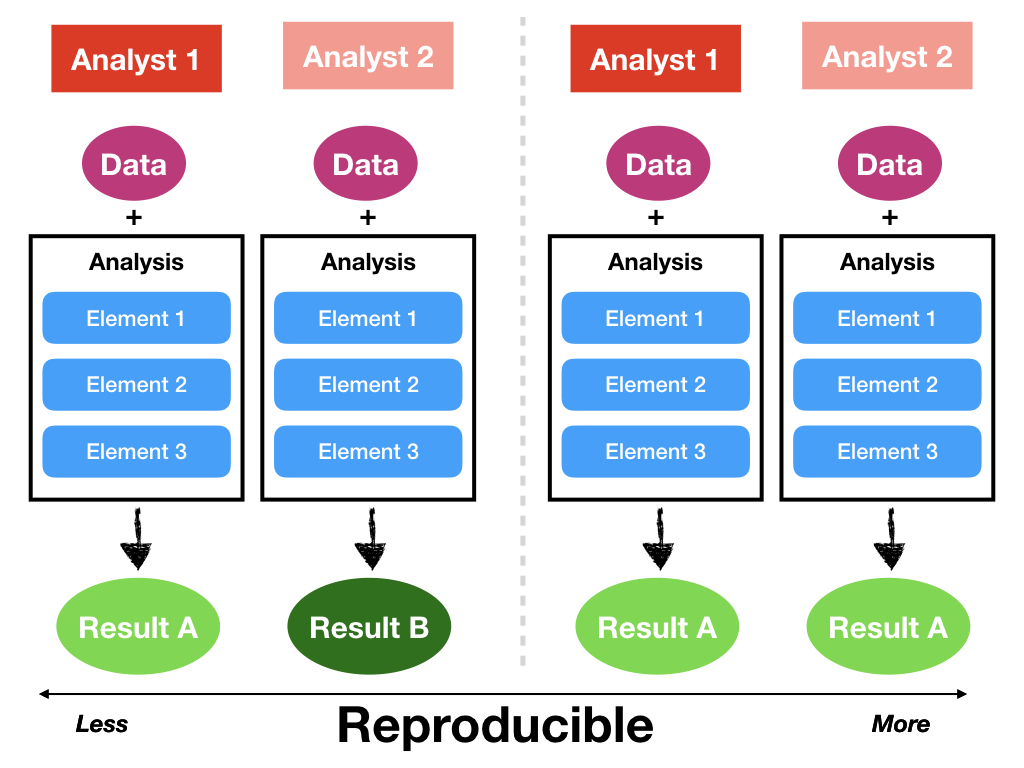}
  \caption{\textbf{The reproducible principle of data analysis.} An analysis is reproducible if someone who is not the original producer (Analyst 2) can take the same data and the same elements of the data analysis and produce the exact same results as the original producer (Analyst 1). In contrast, analyses that conclude in different results are less reproducible.}
  \label{fig-reproducible}
\end{figure}

\end{document}